\documentstyle[11pt,twoside,psfig,jltp]{article}

\begin{document}




\title{Long Range Proximity Effect in Hybrid Ferromagnetic/Superconducting
Nanostructures}


\author{V. T. Petrashov\address{Department of Physics, Royal Holloway,
University of London, Egham, Surrey, TW20 0EX, U.K.}, I. A. Sosnin, C. Troadec}

\runninghead{V. T. Petrashov \it{et al.}}{Long Range Superconducting Proximity Effect}

\begin{abstract}
\par We find that the dependence on temperature and magnetic field of the electrical
resistance of diffusive ferromagnetic nano-wires measured with superconducting electrodes
changes drastically with the distance, $L$, between the ferromagnet/superconductor
contacts, however is remarkably similar for the wires with the same $L$ ranging from 300
nm to 1000 nm, prepared under identical conditions. The result gives an evidence for the
long-range superconductor-induced changes in transport properties of ferromagnetic
nano-wires.

\end{abstract}


\maketitle

\section{Introduction}
\par Recent experimental discoveries of large superconductor-induced changes in the
resistance of disordered ferromagnetic nano-wires suggesting long-range superconducting
proximity effects \cite{1,2,3,4,5,6} have stimulated a significant number of theoretical
investigations \cite{7,8,9,10,11,12,13}, since according to the existing views
superconducting correlations cannot penetrate in the bulk of ferromagnetic materials. One
of the explanations of the effects put forward by authors of [13] is based on the
properties of the $F/S$ interfaces without taking into account proximity induced changes
in the bulk of ferromagnetic conductors. Experiments enabling to separate the bulk and
interface effects in hybrid $F/S$ nanostructures are in order.

\par So far we concentrated on experimental separation of the bulk and interface contributions
using measurements with $F$-wires of different thickness and residual resistance.
Recently \cite{14} we reported new results on the dependence of the proximity effects in
diffusive mesoscopic $F/S$ structures on applied magnetic field. In this paper we focus
on the study of the proximity effects in diffusive $F$-wires of very similar electrical
properties, geometry and crystalline structure however with different distance, $L$,
between $F/S$ contacts. We find that the dependence of the effects on temperature and
magnetic field being drastically different in wires with different $L$, is remarkably
similar for different wires with the same $L$ in the range from 300nm to 1000 nm, giving
a direct experimental evidence for the long-range nature of the proximity effect in
diffusive ferromagnetic wires. We discuss a new mechanism for the long-range effect based
on the analysis of the topologies of actual Fermi-surfaces in ferromagnetic metals.

\section{Experimental}

\par The structures were made using "lift-off" e-beam lithography technique.
The first layer was Ni wire (electrodes 1-7) in contact with Al wires of the second layer
(electrodes 2-6 and 8-12) (see Fig. 1).
\begin{figure}[hb!]
\centerline{\psfig{file=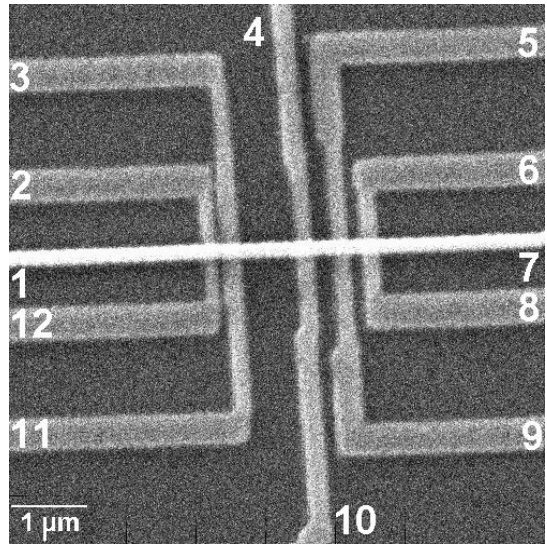,height=3in}} \caption{SEM picture of one of the measured
samples. Contacts 1-7 are Ni 40nm; contacts 2-6 and 8-12 are Al 40nm for one sample and
Al 80nm for the other. Both samples are of exactly the same geometry.} \label{Fig.1}
\end{figure}
We have developed a technique of plasma cleaning of the surface of Ni film followed by
the deposition of Al without breaking vacuum with the resistance of down to 5x10$^{-2}$
$\Omega$ for 100x100 nm$^{2}$ contacts. We measured the resistance of Ni wires with
distance $L$ between $F/S$ contacts in the range from 300 nm to 1000 nm, in the
temperature range from 0.28K to 6K in magnetic fields up to 5T applied perpendicular to
the substrate. All the measured Ni wires were deposited simultaneously and had thickness
of 40 nm, width of 200 nm and the value of $\rho$ of 14 $\mu \Omega$cm corresponding to
the diffusion constant, $D$, of about 40 cm$^{2}$/s. The Al wires were 100 nm wide. Two
batches of samples with Al thickness of 40 nm and 80 nm were prepared under identical
conditions with the deposition of Ni, the spinning of the resist and the baking made
$simultaneously$. The values of $\rho$ and $D$ for Al were 0.8 $\mu \Omega$cm and 200
cm$^{2}$/s, correspondingly. We used commercially available materials of purity
99.999{\%} from Advent Ltd.

\section{Results}
\par
\begin{figure}[hb!]
\centerline{\psfig{file=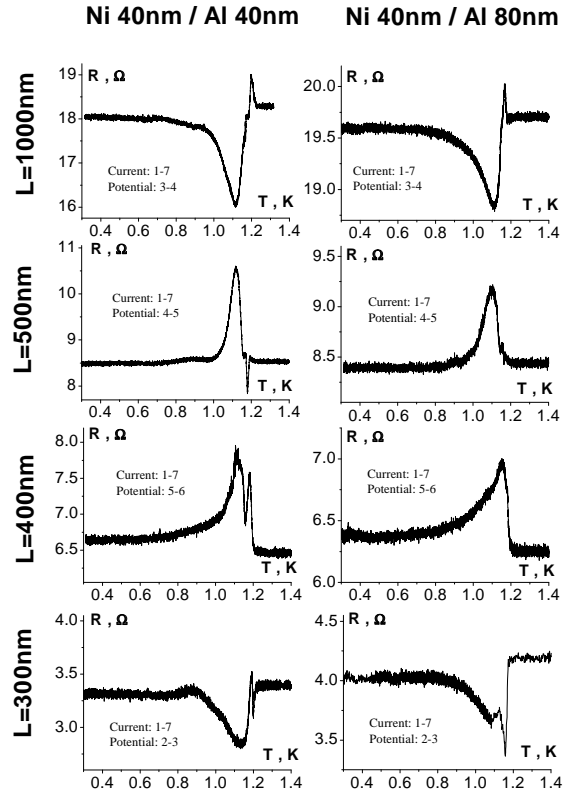,height=4.3in}} \caption{Temperature dependence of the
resistance of the two samples with various length, $L$, between superconducting
electrodes. Current and potential leads are marked according to Fig. 1.} \label{Fig.2}
\end{figure}
The results of measurements of the dependence of the resistance on temperature for 8 Ni
structures are shown in Fig. 2. For all the results presented on Fig. 2 and 3 we used Ni
wires as current leads and Al wires as potential leads. It is seen that the structures
with different distance, $L$, between the $F/S$ contacts show completely different
dependence. Remarkably, the curves for the structures with the same $L$ are similar. The
drop in the resistance reaches -3 $\Omega$ for the sample with $L$=1000nm and 40nm thick
Al that is an order in magnitude larger, than the total contact resistance of 0.18x2
$\Omega$ in the normal state. The amplitude of the effect is larger for samples with 40
nm thick Al probes. That can be accounted for by the higher transparency of $F/S$
interfaces for that sample, where the resistance of contacts in normal
\begin{figure}[hb!]
\centerline{\psfig{file=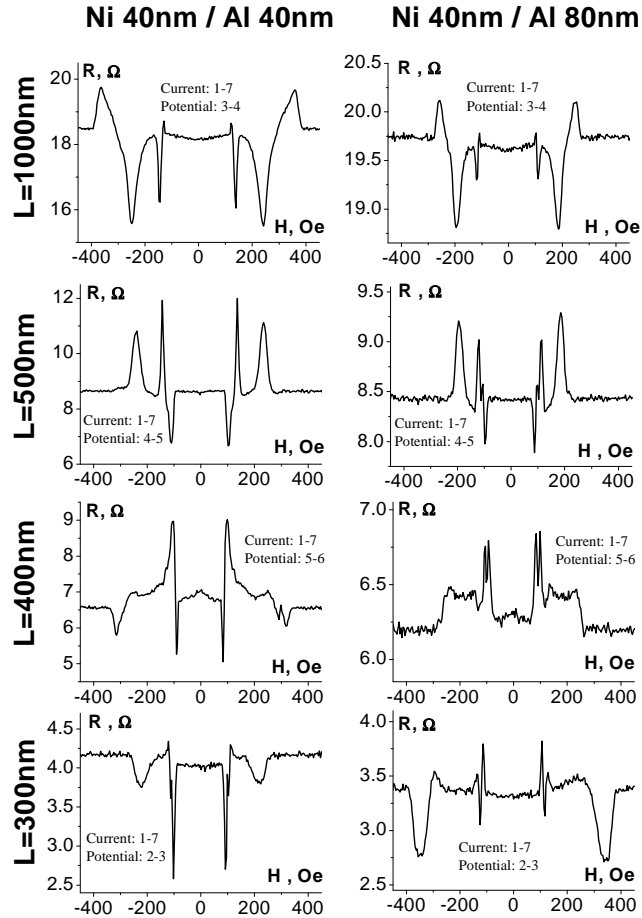,height=5in}} \caption{Magnetoresistance of the
structures of Fig. 2.} \label{Fig.3}
\end{figure}
state was 0.2 ± 0.02 $\Omega$. The voltage-current characteristics of all of the
interfaces from Ni40nm/Al40nm sample, were $N$-shaped in the superconducting state
suggesting strong non-equilibrium effects (Fig. 5 left). The resistance of the contacts
for Ni40nm/Al40nm structures had a scatter in the range from 0.1 $\Omega$ up to 1.0
$\Omega$ with linear $V$ vs. $I$ curves (Fig. 5 right). Nevertheless, an obvious
correlation between the temperature-dependent resistance of samples with the same $L$
persisted despite such a significant difference in the properties of contacts.

\par
\begin{figure}[hb!]
\centerline{\psfig{file=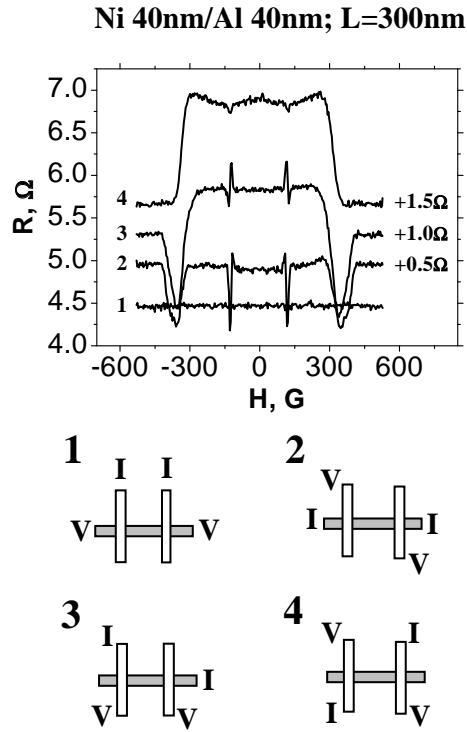,height=4in}} \caption{Magnetoresistance of the Ni
40nm/Al 40nm structure with $L$=300nm measured using different combinations of current
and potential leads. Curves are shifted for clarity.} \label{Fig.4}
\end{figure}
Figure 3 shows magnetoresistance curves for the samples of Fig. 2 measured using the same
combinations of leads at $T=0.28$ K.

The behaviour of the resistance at the onset of superconductivity at critical magnetic
fields correlates with that at critical temperature. The difference between the
resistance in the normal limit at high field and that in zero field coincides with the
difference of the resistance above critical temperature and that at $T=0.28$ K for all
samples, however the dependence on magnetic field is more complicated. Singularities in
the shape of sharp peaks and dips appear in the vicinity of 100 G.

\par We find that the results of measurements depend strongly on the combination of
potential and current leads, as expected. Figure 4 shows the magnetoresistance of one of
the samples measured using different leads. The difference in the measured resistance is
significant. It is much larger than the resistance of the contacts in normal state. The
distribution of the electrochemical potentials in our system depends on the current leads
used with the potential leads of different materials sensing different electrochemical
potentials \cite{15}. We believe that these non-equilibrium effects may result in an
$N$-shaped voltage-current characteristic of the contact seen on Fig. 5 left.

\par
\begin{figure}[ht!]
\centerline{\psfig{file=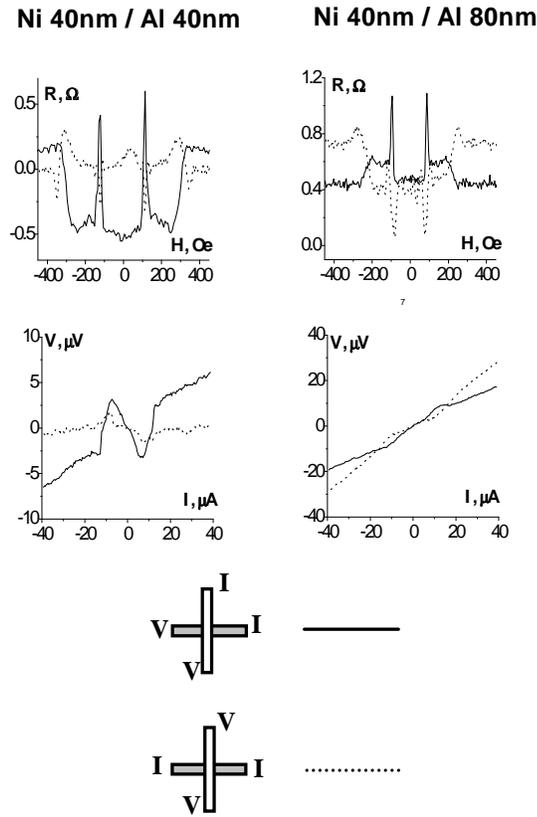,height=4.5in}} \caption{Magnetoresistance and
voltage-current characteristics of one the interfaces from Ni 40nm/Al 40nm sample (left)
and Ni 40nm/Al 80nm (right) sample. Solid and dashed lines correspond to two different
combination of current and potential leads.} \label{Fig.5}
\end{figure}
We have extensively studied the sharp singularities and found that they are the result of
the changes taking place due to magnetic field in the vicinity of $F/S$ contacts. Figure
5 shows the results of measurements of two contacts for two different samples. Note that
there is a voltage drop between the ends of the selfsame Al wire crossing Ni wire. That
means, first, that there are normal regions in Al wire. We believe that Al can go normal
due to reciprocal exchange proximity effect \cite{16} in the vicinity of the $F/S$
contact. Second, there should exist parallel to the Al wire component of current in the
$F/S$ contact. The latter may take place in asymmetric contacts with non-uniform
distribution of the barrier resistance at the $F/S$ interface. Direct measurements (see
Fig. 5) confirm the existence for such an asymmetry. An additional asymmetry is
introduced in the same magnetic fields as sharp singularities in the magnetoresistance.

\section{Discussion}
\par The existing theories are based on idealized isotropic models of the
Fermi-surfaces of ferromagnetic metals. In a ferromagnet with the exchange field energy,
$h_{0}$, the Andreev reflected quasiparticles acquire a momentum of $Q=2h_{0}/v_{F}$ ,
where $v_{F}$ is the Fermi velocity \cite{11}, resulting in an exponential decay of the
superconductor-induced wave functions in diffusive conductors over microscopic distances,
$\xi_{m}=\sqrt{\frac{\hbar D}{2\pi k_{B}T_{0}}}$, where $T_{0}\approx h_{0}/k_{B}$ is the
Curie temperature, $D$ is the diffusion constant. Hence the amplitude of long-range
effects in diffusive $F/S$ systems with small superconducting gap, $\Delta \ll h_{0}$, is
predicted to be negligibly small. We emphasize \cite{17,18,19} that in real metals the
exchange interaction is anisotropic with the value of $Q$ strongly depending on the
position of the Andreev reflected electrons on the Fermi-surface. The value of $Q$ may
$vanish$ for certain electron pairs with opposite spins and directions of momentum (mixed
spin regions at the Fermi surface) electron \cite{20}. Such electrons may be
Andreev-reflected with no effects of the exchange interaction and hence originate the
long-range proximity effects in hybrid $F/S$ nanostructures. The amplitude of the
superconducting condensate functions on the $F$-side should depend on both, the number
the mixed spin electrons and their life times. The latter in principle can be larger than
the averaged over the Fermi surface transport relaxation time (see e.g. \cite{18} and
references therein).
\par One more effect which deserves consideration is the variation
in the exchange interactions at the ferromagnet/vacuum and ferromagnet/substrate
interfaces.

\section{Summary}
\par We summarise the results of the present work that cannot be
accounted by existing theories without assuming long-range proximity effects in
ferromagnetic wires.

\par a) {\it Large drops} in the resistance, $\Delta R$, of up to 3 $\Omega$ at normal state barrier
resistance of 0.2 $\Omega$. The theory predicts $\Delta R < 0.1 \Omega$ for this case.

\par b) {\it Strong correlation} of the proximity induced effects with distance between $F/S$
contacts in the range of up to 1000 nm, even though interfaces themselves were quite
different.

\par c) {\it Strong non-equilibrium effects} with $N$-shaped $V-I$ curves for $F/S$
contacts.

\par d) {\it Sharp singularities} of large amplitude in the magnetoresistance of $F$-wires with
$S$-contacts.

\section{Acknowledgments}

We acknowledge financial support from the EPSRC (Grant Ref: GR/L94611).


\begin{thebibliography}{99}

\bibitem{1}V.T. Petrashov, V.N. Antonov, S.V. Maksimov, and R.Sh. Shaikhaidarov, JETP
Lett. {\bf59}, 551 (1994).
\bibitem{2}M.D. Lawrence and N. Giordano, J. Phys. Cond. Matt. {\bf8}, L563
(1996).
\bibitem{3}M. Giroud, H. Courtois, K. Hasselbach, D. Mailly, and B.
Pannetier, Phys. Rev. B {\bf58}, 11872 (1998).
\bibitem{4} V.T. Petrashov,  I.A. Sosnin, I. Cox, A. Parsons, and C.
Troadec, Phys. Rev. Lett. {\bf83}, 3281 (1999).
\bibitem{5} M. Giroud, K. Hasselbach, H. Courtois, D. Mailly, and B.
Pannetier, in Mesoscopic Superconductors and Hybrid Structures, COST-TMR-CCP9 Workshop,
16-19 December 1999, Lancaster, UK.
\bibitem{6}M.D. Lawrence and N. Giordano, J. Phys. Cond.
Matt. {\bf11}, 1089 (1999).
\bibitem{7} M.J.M. de Jong and C.W.J. Beenakker, Phys. Rev.
Lett. {\bf74}, 1657 (1995).
\bibitem{8}E.A. Demler, G.B. Arnold, and
M.R. Beasley, Phys. Rev. B {\bf55}, 15174 (1997).
\bibitem{9} F.J. Jedema, B.J. van Wees, B.H. Hoving,
A.T. Filip, T.M. Klapwijk, Phys. Rev. B {\bf60}, 16549 (1999).
\bibitem{10}V.I. Fal'ko, C.J. Lambert, A.F. Volkov,
JETP Lett. {\bf69}, 532 (1999).
\bibitem{11}F. Zhou and B. Spivak, preprint, cond-mat/9906177.
\bibitem{12}A.A. Golubov, preprint, cond-mat/9907194.
\bibitem{13}W. Belzig, A. Brataas, Yu.V. Nazarov, G.E.W. Bauer, cond-mat/0005188.
\bibitem{14}V.T. Petrashov,  I.A. Sosnin, I. Cox, A. Parsons, and C.
Troadec, preprint, cond-mat/0005437.
\bibitem{15}V.V. Schmidt, {\it The Physics of Superconductors}, P. Muller and A.V. Ustinov (Eds), Springer, 1997.
\bibitem{16}P.M. Tedrow, J.E. Tkaczyk, A. Kumar, Phys. Rev. Lett. {\bf56}, 1746 (1986).
\bibitem{17}G. Lonzarich in Electrons at the Fermi surface, ed. M. Springford, Cambridge
University Press, 1980.
\bibitem{18}Int. Conf. on Electron life times in metals, J. Phys. Cond. Matt. {\bf19}, 3 (1975); A.K. Geim, V.T.
Petrashov, and M. Zolotarev, Sov. Phys. JETP {\bf91}, 2101 (1986).
\bibitem{19}R. Gersdorf, Phys. Rev. Lett. {\bf40}, 344 (1978).
\bibitem{20}C.S. Wang and J.
Callaway, Phys. Rev. B {\bf15}, 298 (1977).

\end{thebibliography}
\end{document}